\documentclass{jors}
\usepackage[T1]{fontenc}
\usepackage[style=numeric]{biblatex}
\usepackage{graphicx}
\usepackage{booktabs}
\usepackage{textcomp}
\definecolor{mygray}{gray}{0.6}
\definecolor{sarah_edits}{rgb}{0, 0.7, 0.7}

\newcommand{\noun}[1]{\textsc{#1}}

\usepackage{ dsfont }

\usepackage{ulem} 

\begin{document}

{\bf Software paper for submission to the Journal of Open Research Software} \\

To complete this template, please replace the blue text with your own. The paper has three main sections: (1) Overview; (2) Availability; (3) Reuse potential. \\

Please submit the completed paper to: editor.jors@ubiquitypress.com

\rule{\textwidth}{1pt}

\section*{(1) Overview}

\vspace{0.5cm}

\section*{Title}
Mobile EEG artifact correction on limited hardware using artifact subspace reconstruction

\section*{Paper Authors}

1. Maanen, Paul;
2. Debener Stefan;
3. Blum, Sarah

\section*{Paper Author Roles and Affiliations}
1 Neuropsychology Lab, Department of Psychology and Auditory Signal Processing and Hearing Devices, Department of Medical Physics and Acoustics and and Cluster of Excellence "Hearing4all", University of Oldenburg,  Oldenburg, Germany \\
2 Neuropsychology Lab and Cluster of Excellence "Hearing4all", Department of Psychology, University of Oldenburg,  Oldenburg, Germany \\
3 Hörzentrum Oldenburg gGmbH and Neuropsychology Lab and Cluster of Excellence "Hearing4all", Department of Psychology, Oldenburg, Germany

\section*{Abstract}
Biological data like electroencephalography (EEG) are typically contaminated  by unwanted signals, called artifacts. Therefore, many applications dealing with biological data with low signal-to-noise ratio require robust artifact correction. For some applications like brain-computer-interfaces (BCI), the artifact correction needs to be real-time capable.  Artifact subspace reconstruction (ASR) is a statistical method for artifact reduction in EEG. However, in its current implementation, ASR cannot be used in mobile data recordings using limited hardware easily. In this report, we add to the growing field of portable, online signal processing methods by describing an implementation of ASR for limited hardware like single-board computers. We describe the architecture, the process of translating and compiling a Matlab codebase for a research platform, and a set of validation tests using publicly available data sets. The implementation of ASR on limited, portable hardware facilitates the online interpretation of EEG signals acquired outside of the laboratory environment.

\section*{Keywords}
mobile EEG; ASR; artifact-correction; data-processing

\section*{Introduction}
Electroencephalography (EEG) is a non-invasive method for the recording of brain-electrical activity. EEG signals can be recorded from several scalp sites concurrently with small, portable devices. Therefore it is a strong contender for the interface part of BCIs\cite{lotte2014, bottelecocq2014, blankertz2016, liyanage2020, zander2011}. BCIs are systems that allow a direct link between a human and a computer by interpreting brain signals in near real-time and translating these signals to commands for the machine. There is demand for BCIs in the future to be mobile in order to reach their target audience and be useful in everyday situations outside of research contexts \cite{dauwels2016, na2021, ogino2020}. As BCIs have to react to changes in the user's brain state near instantaneously, EEG signal processing for BCIs has to be capable of low-latency, real-time operation, ideally on artifact-free data to avoid the influence of artifacts on the system.\\

Another field where mobile artifact handling is needed is neuropsychology research using recordings of human behavior during motion or natural environments rather than in the laboratory \cite{jacobsen2020, debener2012, devos2014, devos2014b} . This implicates a need for mobile research platforms capable of recording and processing EEG signals, including before applying specific analyses steps. \\

However, both BCI and mobile EEG signals are prone to be contaminated by artifacts, in movement even more than stationary data \cite{jacobsen2020}. Artifact correction therefore is indispensable for mobile EEG signal processing due to the fact that many abnormal data occur irregularly and with high amplitudes and can thus hinder correct interpretation of the data\cite{chandola2009, blum2019, mane2020, chaudhary2021, alchalabi2021}. Artifacts can have a large influence on the performance of signal processing methods, they may especially break the assumption of some methods that certain statistical properties remain stable in the data. Artifact correction for BCIs of the future and for mobile EEG experiments needs to be capable of online operation on mobile hardware and needs be able to handle artifacts that occur in mobile EEG. \\

One candidate method is artifact subspace reconstruction (ASR) \cite{mullen2015}, a statistical method for artifact reduction in EEG. It has been shown to perform well with movement artifacts and eye blinks \cite{blum2019, jacobsen2020, ladouce2021, dehais2020, plechawska2018, nathan2016, nordin2018, dehais2020}, artifact classes which are especially important for mobile EEG.\\
\paragraph*{The artifact subspace reconstruction algorithm}
ASR is a statistical artifact correction method \cite{mullen2015, blum2019, chang2018, pion2018}. It detects artifacts based on their abnormal statistical properties when compared to artifact-free data. After detection, a correction is applied and the result of the method are data with the same amount of samples and channels as the (possibly corrupted) input data.
To achieve this correction, ASR defines a statistical subspace based on presumably artifact-free calibration data in an initial calibration phase. The statistical subspace is defined by the distribution of the calibration data. It is computed based on channel covariance matrices of calibration data, exploiting the fact that EEG channels correlate highly with their direct neighbourhood \cite{mullen2015, blum2019, chang2018}.
After calibration, the subsequent processing phase consists of a comparison of incoming data chunks to the properties of these artifact-free calibration data using a similarity measure based on distance in the statistical subspace. If a given chunk of data is detected to contain artifactual influences, a correction is applied to the affected segment.
The correction is applied in the statistical subspace as well and is performed using eigenvectors of covariance matrices of the calibration data.

ASR benefits from channel configurations that cover a large area due to the way most artifacts are recorded on many channels at the same time \cite{blum2019, chandola2009, islam2016}). Nevertheless, the computational complexity of the method is relatively low, even with large channel numbers. The reasons lie in the representation of the multi-channel data in covariance matrices which remain of low dimensionality while preserving relevant statistical properties such as covariance among neighboring channels.

However, because ASR is implemented in Matlab, in its current implementation, thereafter called \textit{reference implementation}, ASR cannot be used in mobile EEG recordings easily. Matlab needs a decently powerful PC\cite{matlab-requirements} and is only available for operating systems running on an Intel-architecture processor. Mobile recording hardware is typically not fast enough to run Matlab and usually sports some variation of ARM architecture.
There have been efforts to modify ASR for mobile devices \cite{van2021}, but they aim mainly at field programmable gate arrays (FPGAs), which are more specialised and harder to use for the average scientist, and there is no open source implementation of mobile ASR as of yet.
\\

Therefore we propose mobile ASR (mASR), an implementation of ASR for limited hardware, in our case a research platform to evaluate audio/time domain/hearing aid signal processing. In this report, we describe the architecture, the process of translating and compiling a Matlab codebase for this research platform, and a set of validation tests using publicly available data sets. With the implementation of ASR on portable hardware, the interpretation of neural data in different contexts is possible. Furthermore, we hope that this report provides a guideline for others looking to translate Matlab code into compiled languages for usage on limited hardware. \\

Our target hardware is the Portable Hearing Laboratory (PHL) \cite{pavlovic2018, phl-webpage}, a mobile integrated setup for audiological research based on the \noun{Beaglebone Black Wireless} single-board computer (see tab \ref{tab:phl-specs} for hardware specs) and the open-source audio interface \noun{Cape4All}\cite{herzke2018}. It includes a battery and a set of binaural behind-the-ear (BTE) hearing aids, and runs a software platform for various audio signal processing algorithms, openMHA \cite{kayser2022}, on a Linux distribution optimized for low-latency real-time audio processing. The PHL has been and is being used for mobile audiological experiments in the past \cite{dasenbrock2021, luengen2021, denk2021}. There are currently efforts ongoing to integrate EEG sensor data into the the openMHA audio processing, fueled by the rising need of smarter hearing aids that adapt the audio processing based on context information \cite{dasenbrock2021, aroudi2020, noymai2019}. Its established user base together with the urgent need for EEG artifact reduction makes the PHL an attractive target platform for mASR. \\
\begin{table}
\begin{tabular}{lr}
\toprule 
CPU & Octavo Systems OSD3358 1GHz ARM® Cortex-A8\tabularnewline
\midrule 
RAM & 512MB DDR3 RAM\tabularnewline
\midrule 
Size & ca $8.5\,\mathrm{cm}\times6\,\mathrm{cm\times3\,\mathrm{cm}}$\tabularnewline
\midrule 
Weight & $135\,\mathrm{g}$\tabularnewline
\midrule 
Battery life & $4-6+\,\mathrm{hours}$, depending on load\tabularnewline
\bottomrule
\end{tabular}
\protect\caption{PHL specs\label{tab:phl-specs}}
\end{table}

\paragraph*{openMHA in a nutshell}
Since openMHA is described in detail elsewhere \cite{openmha-doc, kayser2022} we only introduce some concepts especially important for our plugin. The openMHA main executable acts as a plugin host while the plugins provide the actual signal processing. Every plugin needs to implement the \texttt{prepare()}, \texttt{process()} and \texttt{release()} functions. The \texttt{prepare()} method is called before the processing starts. It locks in the input signal dimensions and provides the framework with the output signal dimensions. \texttt{process()} is the main processing method. It receives the input signal and returns the output signal and is called periodically by the framework. The \texttt{release()} method is used to free resources after processing. The openMHA framework knows different kinds of variables: \noun{Algorithm communication variables} (AC variables) are used to share side-channel information between plugins while \noun{configuration variables} are used to configure plugins. openMHA provides a mechanism to change configuration variables during audio processing in a real-time safe manner. 

\section*{Implementation and architecture}
\begin{figure}
\begin{centering}
\includegraphics{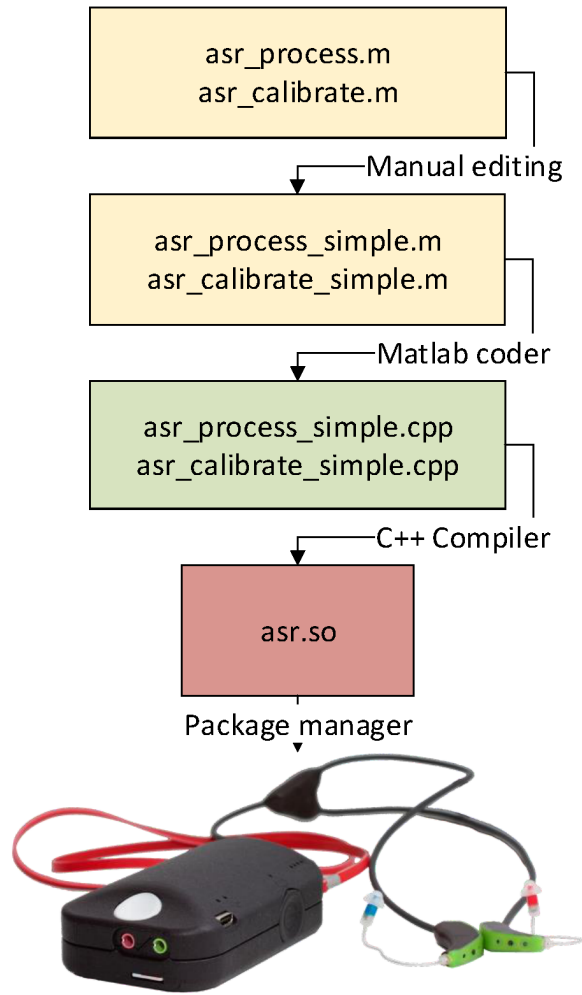}
\par\end{centering}
\caption{The mASR build pipeline. Ochre stands for Matlab code, green for C++
Code and red for compiled code\label{fig:pipeline}. PHL image taken from \cite{mha-hardware-webpage}. For PHL specs see table \ref{tab:phl-specs}. The PHL is available from BatAndCat Labs, Palo Alto, USA.}

\end{figure}

In order to reduce development time the Matlab Coder was used to automatically translate a simplified version of the reference implementation to C++. 
In order to be able to use the automatic translation from Matlab to C or C++ code, some adaptations to the original implementation needed to be made. They are described later in \textit{Translation of ASR to Matlab Coder-compatible source base}.
Apart from the simplification, the translation process was automated by using Makefiles, thereby shortening the time it takes to integrate improvements in the reference implementation and reducing the potential for human error. Figure \ref{fig:pipeline} shows a sketch of the build pipeline.
\paragraph*{mASR in the openMHA framework}
Our implementation consists of several files, of which the most important are:
\begin{description}
\item [{codegen}] contains the C++ code generated by the Matlab coder, including the so-called entry-point functions \texttt{asr\_calibrate} and \texttt{asr\_process}, which are the equivalent of the Matlab functions of the same name.  
\item [{Makefile}] The Makefile providing build automation.
\item [{asr\_calibrate\_simple.m}] The simplified version of the asr\_calibrate.m script.
\item [{asr\_process\_simple.m}] is the simplified version of asr\_process.m
\item [{make.m}] steers the Matlab Coder configuration.
\item [{asr.cpp}] contains the implementation of the openMHA plugin
\end{description}
Our plugin implementation follows the usual structure of openMHA plugins and implements the usual openMHA callback functions. The incoming EEG signal is provided via AC variable, the usual inter-plugin communication method in openMHA. The \texttt{stride} attribute of the AC variable represents the number of channels in the EEG signal, which has the same requirements as the input signal of the reference implementation. To import the signal into openMHA, other plugins have to be used, for example the \texttt{lsl2ac}-plugin, which converts LSL\cite{stenner2021} streams into AC variables.
The most important functions are the \texttt{prepare()}, the \texttt{process()}, and the \texttt{release()} functions.
The \texttt{prepare()} function is where most initialization takes place and the configuration variables are locked in. It also performs the calibration. The calibration is provided as csv file, where every row presents a channel.
\texttt{process()} is the main processing function. It is called periodically by the audio processing thread of the openMHA executable. To ensure real-time safety, the actual artifact reduction happens in a separate thread. The input data is shared between the threads safely by means of a lock-free first-in, first-out (FIFO) buffer. The output of the cleaning algorithm is then shared with the processing thread via another FIFO buffer and made accessible to the downstream plugins as an AC variable. The output AC variable can then processed within openMHA or be exported and processed elsewhere, e.g. via the \texttt{ac2lsl} plugin.
\texttt{release()} signals the ASR thread to terminate and unlocks the configuration variables. The plugin has the following configuration variables:
\begin{description}
    \item [{SamplingRate}] the nominal sampling rate of the input data in Hz.
    \item [{WindowLength}] Length of the statistics window, in seconds (e.g., 0.5). This should not be much longer than the time scale over which artifacts persist, but the number of sample in the window should not be smaller than 1.5x the number of channels.
    \item [{VarName}] Name of the input AC variable.
    \item [{CalibrationFileName}] file name of the calibration file.
\end{description}

\paragraph*{Translation of ASR to Matlab Coder-compatible source base}
In order to be able to generate C/C++ code from the ASR Matlab codebase, some steps with regards to compatibility to the Matlab code generation needed to be implemented. 
Firstly, all unsupported functions were replaced (complete list of supported functions: \url{https://de.mathworks.com/help/referencelist.html}.
This included the unrolling of the \texttt{bsxfun} function into explicit loops (\url{https://de.mathworks.com/help/matlab/ref/bsxfun.html}). \texttt{bsxfun} allows to apply element-wise operations by providing a function and an array, \texttt{bsxfun} implicitly expands the application to all elements and the resulting operation can be equivalently expressed by looping through the array. Another example is slicing of vectors or array sections, which the Matlab Coder cannot expand to all datatypes automatically. Instead, an explicit indexing was implemented at the corresponding code locations.  
Then, all variables were declared with their desired size and initialized before their first usage in both \texttt{asr\_calibrate.m} and \texttt{asr\_process.m}.
Subsequently tests were performed to ensure the equivalency of the adapted ASR code to the original code base.

\section*{Quality control}
In order to ensure that mASR works correctly, we processed data sets using the reference implementation and our implementation and then compared the results of \texttt{asr\_calibrate\_simple()} and its C++-counterpart and \texttt{asr\_process\_simple()} and its C++-counterpart. We deem the test for \texttt{asr\_process\_simple()} passed when all outputs of the calibration and processing have the same value as the output of the reference implementation, up to a relative tolerance of $10^-5$. Test data adapted from a publicly available data set \cite{holtze2021} are included in the mASR distribution on GitHub. In order to run the tests, the user needs to run 
\begin{verbatim}
make test; ./test
\end{verbatim}
within the main repository directory. 
\section*{(2) Availability}
\vspace{0.5cm}
\section*{Operating system}
mASR is tested using Matlab R2021b and openMHA 4.16. It can run with operating systems for which these are available, including Windows 7 or later, macOS and Linux.
\section*{Programming language}
Matlab, C++
\section*{Additional system requirements}
The translation from Matlab to C++ code needs a PC capable of running Matlab. The compiled code should run anywhere a Linux operating system is available, including very limited hardware. 
\section*{Dependencies}
Matlab R2021b: Matlab Coder 5.1 \\
openMHA 4.16. \\
Optional: \\
GoogleTest \\
GoogleBenchmark

\section*{List of contributors}
\begin{itemize}
\item Paul Maanen (Oldenburg University) Developed the plugin and set up the build pipeline
\item Sarah Blum (Hörzentrum Oldenburg) Adapted the reference implementation for Matlab Coder
\item Stefan Debener (Oldenburg University) Provided the lab and gave feedback on the manuscript
\end{itemize}

\section*{Software location:}

{\bf Archive}
\begin{description}[noitemsep,topsep=0pt]
	\item[Name:] Zenodo
	\item[Persistent identifier:] https://doi.org/10.5281/zenodo.6362843
	\item[Licence:] GNU AGPL v3
	\item[Publisher:] Paul Maanen
	\item[Version published:] 0.0.3
	\item[Date published:] 16/03/22
\end{description}

{\bf Code repository}

\begin{description}[noitemsep,topsep=0pt]
	\item[Name:] GitHub
	\item[Persistent identifier:] https://github.com/NeuropsyOL/mASR
	\item[Licence:] GNU AGPL v3
	\item[Date published:] 11/02/22
\end{description}

\section*{Language}
English

\section*{(3) Reuse potential}
Our implementation of mobile ASR is useful for all researchers looking for online artifact correction of EEG signals in a mobile package. While we chose to integrate our implementation of ASR into an openMHA plugin, the source code is freely available and extensively documented, making it easy for the end user to modify for other target platforms. We provide support on a best-effort basis via GitHub. Users can request features or report bugs by opening an issue on the GitHub repository.
\section*{Acknowledgements}
We like to thank the original authors and current maintainers of the Matlab ASR implementation.
\section*{Funding statement}
This work was supported by the DFG Cluster of Excellence EXC 1077/1 "Hearing4all".
\section*{Competing interests}
The authors declare that they have no competing interests.

\normalem
\printbibliography
\vspace{2cm}

\rule{\textwidth}{1pt}

{ \bf Copyright Notice} \\
Authors who publish with this journal agree to the following terms: \\

Authors retain copyright and grant the journal right of first publication with the work simultaneously licensed under a  \href{http://creativecommons.org/licenses/by/3.0/}{Creative Commons Attribution License} that allows others to share the work with an acknowledgement of the work's authorship and initial publication in this journal. \\

Authors are able to enter into separate, additional contractual arrangements for the non-exclusive distribution of the journal's published version of the work (e.g., post it to an institutional repository or publish it in a book), with an acknowledgement of its initial publication in this journal. \\

By submitting this paper you agree to the terms of this Copyright Notice, which will apply to this submission if and when it is published by this journal.

\end{document}